\documentclass[preprint,12pt]{elsarticle}
\pdfoutput=1

\usepackage{amssymb}

\usepackage{eurosym}
\journal{NIM A}
\newcommand{\trigsim}{{\tt trigsim}}
\newcommand{\trigsimp}{{\em trigsim}}
\newcommand{\simtel}{{\tt sim\_telarray}}
\newcommand{\eur}{\euro}

\newif\ifjournal

\begin{document}
\begin{frontmatter}


\title{A versatile digital camera trigger for telescopes in the
Cherenkov Telescope Array
}
\author[hu]{U.~Schwanke\corref{cor1}}
\cortext[cor1]{Corresponding author (schwanke@physik.hu-berlin.de)}
\author[desy]{M.~Shayduk\corref{cor2}}
\cortext[cor2]{Corresponding author (maxim.shayduk@desy.de)}
\author[desy]{K.-H.~Sulanke}
\author[desy,hu]{S.~Vorobiov}
\author[desy]{R.~Wischnewski}
\address[hu]{Humboldt--Universit\"at zu Berlin, Newtonstra{\ss}e 15, 12489 Berlin, Germany}
\address[desy]{DESY Zeuthen, Platanenallee 6, 15738 Zeuthen, Germany}

\begin{abstract}
This paper describes the concept of an FPGA-based digital camera trigger for 
imaging atmospheric Cherenkov telescopes, developed for the future 
Cherenkov Telescope Array (CTA). The proposed camera trigger is designed to select 
images initiated by the Cherenkov emission of extended air showers from 
very-high energy (VHE, $E>20$\,GeV) photons and charged particles while suppressing 
signatures from background light. The trigger comprises three stages. A first stage 
employs programmable discriminators to digitize the signals arriving from the
camera channels (pixels). At the second stage, a grid of low-cost FPGAs is used to process the 
digitized signals for camera regions with 37 pixels. 
At the third stage, trigger conditions found independently in any of the overlapping 37-pixel regions are 
combined into a global camera trigger by few central FPGAs. 
Trigger prototype boards based on Xilinx FPGAs have been designed, built and
tested and were shown to function properly. Using these components a full camera 
trigger with a power consumption and price per channel of about 0.5\,W 
and 19~\euro, respectively, can be built. With the described design the camera trigger algorithm can
take advantage of pixel information in both the space and the time domain allowing,
for example, the creation of triggers sensitive to the time-gradient of a
shower image; the time information could also be exploited to online adjust the time window
of the acquisition system for pixel data. Combining the results of the parallel execution of different 
trigger algorithms (optimized, for example, for the lowest and highest energies, respectively) on each FPGA
can result in a better response over all photons energies (as demonstrated by
Monte Carlo simulation in this work). 
\end{abstract}


\end{frontmatter}

\section{Introduction}
\label{sec:intro}

The proposed Cherenkov Telescope Array (CTA, \cite{CtaConceptShort})
is a large installation of Cherenkov telescopes of different sizes for 
the detection of very high-energy (VHE, $E>20$\,GeV) $\gamma$-rays. CTA will cover the energy
range from few tens of GeV up to hundreds of TeV with a sensitivity at 1\,TeV that
is a factor of 10 better than achieved by the current-generation experiments
H.E.S.S.\footnote{http://www.mpi-hd.mpg.de/hfm/HESS}, MAGIC\footnote{http://wwwmagic.mpp.mpg.de}, and VERITAS\footnote{http://veritas.sao.arizona.edu}.
It will also provide a good energy (about 10--15\,\%) and angular resolution (on the
arcmin scale) for reconstructed photons. 

Currently considered array designs (see, for example, \cite{CtaMc}) 
deploy a mixture of large-size telescopes (LSTs),
medium-size telescopes (MSTs), and small-size telescopes (SSTs), with typical
reflector diameters of 23\,m, 12\,m, and 4\,m, respectively, on
an area of roughly 1--10\,km$^2$ in order to ensure good performance
over four orders of magnitude in photon energy. The enlarged
instrumented area and telescope field of view (FOV) along with the 
wider enegry range (when compared to current-generation
experiments) imply a particular challenge for the trigger and
and data-acquisition systems which have to deal with a 
cosmic-ray-induced array trigger rate of O(10\,kHz), typically an order of magnitude higher 
than in current installations.

CTA trigger designs typically comprise two trigger levels. 
At the telescope level,
the proposed Cherenkov cameras (equipped with 1,000--10,000 
pixels with diameters corresponding to about $0.3^{\circ}$--$0.1^{\circ}$) are expected to provide local
camera triggers for $\gamma$-ray and cosmic-ray showers while efficiently suppressing 
the background. Typical $\gamma$-rays generate a Cherenkov light-flash of a 
few nanosecond duration in spatially neighbouring 
pixels, but at high energies ($\gg 1\,$TeV) and large impact 
distances of the shower with respect to the telescope the 
camera image acquires a substantial time-gradient (\cite{timegradient}) 
and can last several 10\,ns. The background is dominated by
the diffuse night-sky background (NSB) light from
natural and articial light sources, resulting in pixel
count rates of O(100\,MHz) at single photoelectron (p.e.) 
threshold, and by large-amplitude afterpulses mimicking 
Cherenkov signals in a pixel. Further suppression can
be gained at the inter-telescope level where triggers can
combine the information from spatially neighbouring telescopes or even from
all telescopes in the array. Array-level or inter-telescope 
triggers typically require a coincidence of at least two telescopes in a time
window of several 10\,ns duration \cite{HessArrayTrigger2004,German2008}
or make sure that the camera images are compatible with
the origin from a $\gamma$-ray shower \cite{Krennrich2009}.
Such triggers reject, in particular, NSB triggers and events where a single 
muon from a hadronic shower hit a telescope and generated 
a camera trigger due to its Cherenkov emission.

At the telescope level, a versatile camera trigger is needed
to select $\gamma$-rays over the full targeted energy range
with good efficiency. Ideally, the trigger hardware should
be applicable largely independent of the telescope type and 
the trigger should also provide guidance to the camera-readout
system how much of the image should be kept (both in space
and, in particular in the presence of a large time-gradient
in the image, in time \cite{colibri}). This paper describes the concept of a
digital camera trigger based on Field Programmable Gate Arrays (FPGAs).
Section \ref{sec:digi} 
presents the design and a possible hardware
implementation of the trigger. Section \ref{sec:prototype} 
discusses test results with prototype trigger boards. 
The impact of the proper choice of the camera trigger algorithms
is illustrated in Section~\ref{sec:simu} using the 
results of a Monte Carlo simulation with the \trigsim\ program (described in the appendix).
Conclusions are presented in Section~\ref{sec:summary}.

\section{A Digital Camera Trigger for CTA}
\label{sec:digi}
\subsection{Camera Trigger Strategies}

Camera trigger strategies employed in current-generation
experiments \cite{MagicTrigger2004,zitzer2013} are either based on the topological distribution 
of pixel hits (i.e.~pixels with a signal above a discriminator 
threshold) or on the analogue sum of pixel signals. In the first 
approach (referred to as {\em majority trigger}) one
requires at least $N_{\mathrm{maj}}$ (for example 3) pixels above a certain threshold
(a few p.e.) in a coincidence window of few nanosecond length.
Such trigger designs differ in the definition of pixel groups 
(out of which $N_{\mathrm{maj}}$ pixels must be above threshold) and 
in the required hit pattern (pixels neighbouring or not). 
In the second approach (the {\em sum trigger}), the analogue sum of
all signals in a pixel group (trigger patch) must be greater than a value
${\mathrm{DT_{sum}}}$ (around 20 p.e.). To suppress the impact of
afterpulses the pixel signals are often clipped (for example at 6 p.e.)
before the summation. For both majority and sum triggers
the used pixel groups are usually overlapping in space to avoid losses
of shower images that occur at the boundaries of two pixel groups.

The implementation of majority triggers can be done in an analogue
or digital fashion alike, and also mixed concepts (for example
the fast analogue summation of comparator output signals in a pixel group \cite{HessArrayTrigger2004,HessTrigger2003}) 
have been used. The implementation of an analogue sum trigger
was instrumental in lowering the trigger threshold for pulsar
studies \cite{MagicSumTrigger2008}, but it is clear that also an approximation of such a trigger
can be built using digital electronics. For CTA, with its increased 
number of telescopes and different Cherenkov
camera types, aspects of the camera trigger like costs per channel, 
robustness, adaptability to the camera geometry and the energy
range targeted with a certain telescope type are particularly important.
In this sense, fully digital camera triggers based, for example,
on fast, freely reprogrammable FPGAs may offer advantages over analogue
solutions where the trigger logics has been hardwired. The algorithm
executed in such a trigger scheme can be modified and adapted easily,
and it is also possible to run several trigger
algorithms in parallel and to combine their results to obtain a higher photon efficiency.

For definiteness, the case of an MST with about 2000 
photomultiplier tube (PMT) pixels that are arranged in a specific geometry
will be considered in the following. It is clear that the general
trigger concept can be adapted to other telescopes sizes and 
photon detection technologies (e.g.~silicon photomultipliers or 
multi-anode PMTs).

\subsection{The FPGA Trigger}

The digital FPGA trigger described here is based on the idea
to generate digital camera images with a depth of e.g.~1 bit at a rate of e.g.~1\,GHz 
and to process the images with one type of rather inexpensive FPGA which can
look for pixel coincidences in time and space.
Full image coverage is ensured by the 
processing of overlapping camera regions. The envisaged trigger scheme comprises three levels, L0 to L2:
\begin{description}
\item The {\bf L0 stage} (Section \ref{sec:l0hardware}) imposes a basic signal threshold and digitizes 
    the preamplified PMT signals with the help of a programmable comparator.
    With some further signal processing, the length of the digital signal can 
    be used to encode the time over threshold (TOT)
    of the PMT pulse or an estimate of the signal amplitude derived from the TOT.
\item The {\bf L1 stage} (Section \ref{sec:l1hardware}) consists of FPGAs each of which receives the L0 signals 
    from a camera region that is large enough to contain a good fraction
    of a possible shower image (at most 49 pixels). Overlap of the camera regions
    is ensured by an exchange of L0 signals with neighbouring FPGAs.
    Each FPGA executes freely programmable trigger alorithms in time slices
    of about 1\,ns length and generates a L1 trigger signal for its camera region.   
\item The {\bf L2 stage} (Sections \ref{sec:l2hardware0} and \ref{sec:l2hardware1}) combines the L1 trigger signals from all 
    overlapping camera regions and generates a camera trigger.
\end{description}

\subsection{Trigger Architecture}

\begin{figure}[t]
\includegraphics[width=\linewidth]{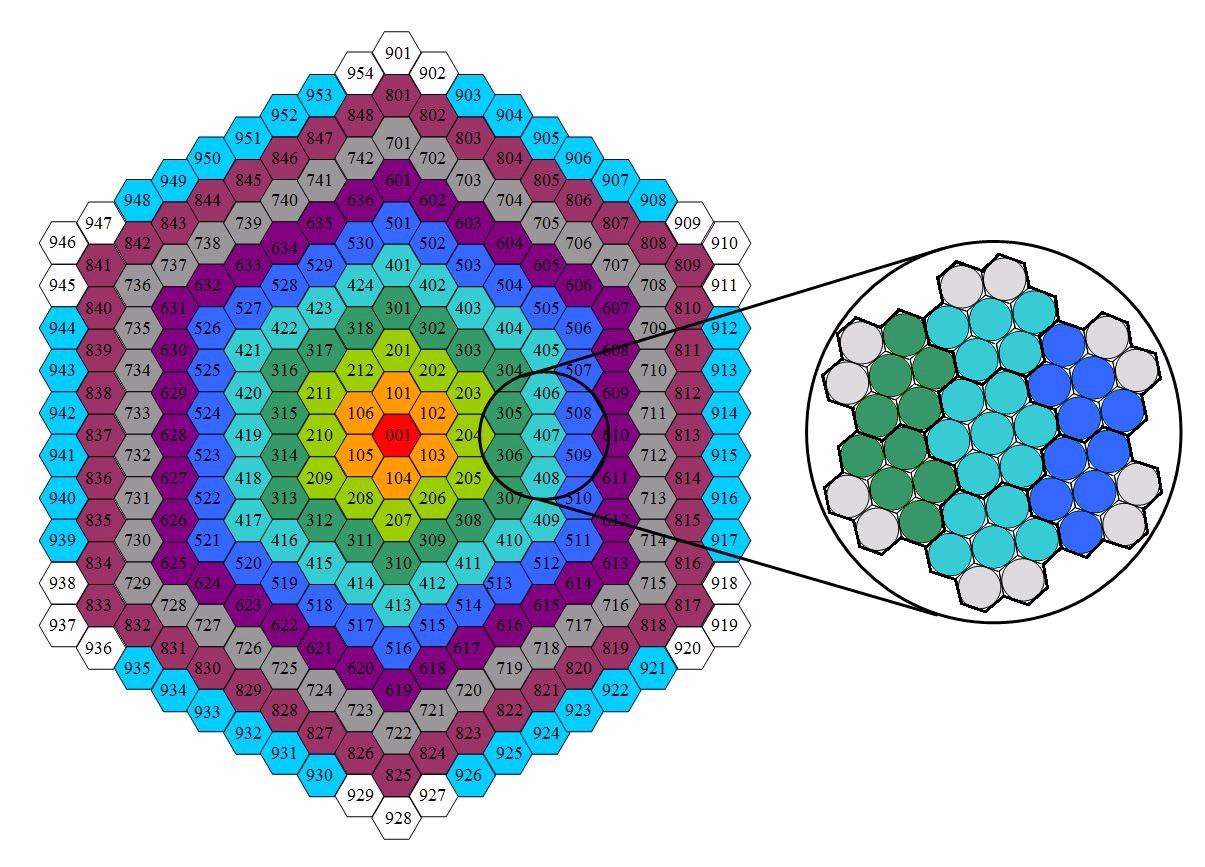} 
\caption{Geometry of a simulated CTA camera. Each hexagon is a so-called
cluster comprising seven PMT pixels. There are 271 clusters comprising a total of
1897 pixels. In the digital trigger scheme, all channels
of a cluster are connected to a cluster FPGA which also receives
L0 signals from its directly neighbouring clusters. The resulting 
super-clusters overlap, and each cluster FPGA executes a trigger algorithm 
in its super-cluster. The inset on the right-hand side shows the 
structure of one selected super-cluster (small black circle) consisting 
of seven 7-pixel clusters. The L0 signals of all pixels except the 
ones shaded with grey colour are processed by the cluster FPGA of the central cluster. 
The total number of L0 signals is 37: seven signals from the central cluster and five signals from each
of the six surrounding clusters.
}
\label{fig:CameraClusters}
\end{figure}

The computing power of low-cost FPGAs, the number of allowed input/output (I/O) channels
and the speed of the links between PMTs and FPGA must be balanced with the size and the required overlap of
camera regions. Some CTA camera designs define a group of 7 pixels (one central pixel
and the six surrounding pixels) as basic building block that can be handled and
exchanged independently of other pixels in a camera. Besides the PMT pixels 
such a so-called {\em cluster} contains also the needed infrastructure (high and low
voltages), front-end electronics (preamplifiers, data buffers), and trigger boards.
Each cluster has six direct neighbours, and a complete CTA camera can be built up by combining some hundred clusters as
illustrated in Fig.~\ref{fig:CameraClusters} for an MST camera with 1897 pixels in 271 clusters. 
An arrangement of
seven clusters (49 pixels) is referred to as {\em super-cluster} and subtends
an angle of O($1^{\circ}$), i.e.~it covers a good fraction of even the largest
shower images. One cluster FPGA is assigned to each cluster and can receive up to 7 L0 signals from each of the 
6 surrounding clusters by means of fast serial links. At the same time up to 7 L0 signals can be transmitted to 
each of the 6 surrounding clusters. Every cluster FPGA is thus the central engine of a
super-cluster. In the design described here, there are 8 LVDS input and 8 LVDS output connections
between any two cluster FPGAs. Each cluster FGPA utilizes 7 inputs from the local cluster and 5
of the 7 inputs from each of the six neighbouring clusters, cf.~the inset in Fig.~\ref{fig:CameraClusters},
i.e.~the presently implemented firmware exchanges only 5 L0 signals with the surrounding clusters.
Every cluster FPGA has thus 37 inputs, except for clusters that are located at the camera boundary and have therefore
fewer inputs. This mapping allows the execution of trigger algorithms on the
central 37 pixels of a super-cluster and creates a sufficient overlap of the super-clusters.
The unused $3\times 2$ fast LVDS channels per connection extend the flexibility of the firmware design.
They can be used to distribute other signals (like the camera trigger signal or a pulse per second (PPS)),
across the camera backplane in a daisy chain way.

Due to the overlap of super-clusters a trigger condition can be found
by more than one cluster FPGA. The detected trigger (the L1 signal) 
is therefore forwarded from all cluster FPGAs to a central FPGA. This FPGA
can execute higher-level algorithms or even estimate image parameters. In the simplest
case, it can derive a main trigger by just performing a logical OR on the triggers delivered by the cluster FPGAs. 
It can also interact with the camera data acquisition system to guide the camera readout
since it has superior knowledge of the time development of the camera image 
possibly covering several super-clusters.

\subsection{Hardware Implementation}

\begin{figure}[t]
\centering%
\includegraphics[width=\linewidth]{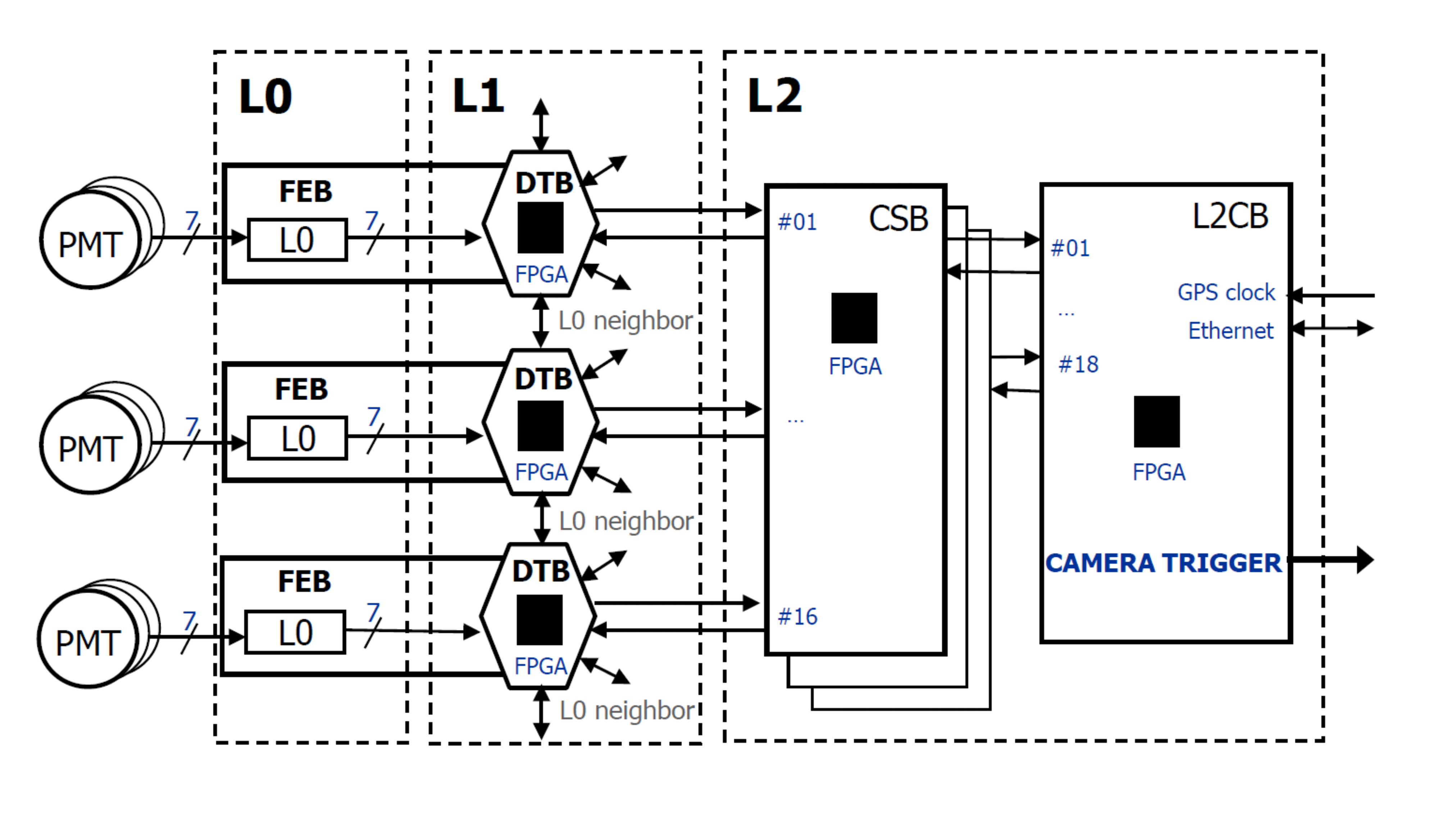}
\caption{
Hardware architecture of the digital trigger. The L0 signals
for a cluster of 7 PMT pixels are generated on the front-end board (FEB) 
and sent to a cluster FPGA on the Digital Trigger Backplane (DTB) board.
Each cluster FPGA exchanges L0 signals with the FPGAs assigned to neighbouring 
clusters ({\em L0 neighbour}) and executes a trigger algorithm on 
37 channels to generate the L1 signal for its super-cluster. The design
for the L2 stage foresees 18 Cluster Service Boards (CSB) that 
accept L1 signals from 16 super-clusters; the total number of pixels
is then 2016. The camera trigger is finally generated by one L2 Controller Board 
(L2CB) based on the input from the 18 CSBs. 
}
\label{fig:TrigHardware}
\end{figure}

Hardware implementations of digital trigger components have been manufactured
as a proof of principle for the concept using Altera\footnote{http://www.altera.com} 
and Xilinx\footnote{http://www.xilinx.com} FPGAs.
In the following, a solution that encodes the TOT in the L0 signal length
and that is based on the Xilinx Spartan 6 FPGA will be described in more detail. 
Figure~\ref{fig:TrigHardware} provides an overview of the different
hardware components that are needed to generate and process the L0, L1 and L2 
signals in a setup with only three 7-pixel clusters. 

The implementation assumes that the L0 signals for seven PMTs are generated on the
front-end board (FEB) that provides the basic infrastructure (voltage supplies, preamplifiers,
data buffers) for a cluster. On the FEB, the PMT signals are
preamplified and passed to a fast Low Voltage Differential Signaling (LVDS) comparator which compares the analogue signal with 
the signal at a second input whose signal level can be adjusted with the help of a 
digital-to-analogue converter (DAC)
and that functions as a threshold. 
If the PMT signal exceeds the threshold, a digital LVDS L0 signal
is created whose pulse length corresponds to the TOT of the PMT signal.
The minimum detectable input pulse width is about 1\,ns full width half maximum (FWHM), the
minimum amplitude is about 0.3 photoelectrons.

A so called digital trigger backplane (DTB) board is mounted behind each cluster. 
The central element of the DTB board is the cluster FPGA receiving the L0 signals from its 
FEB. The FPGA exchanges L0 signals ({\em L0 neighbour}) with up to six neighbouring DTB
boards by means of fast serial links and executes a trigger algorithm on the 37 channels
assigned to it to generate a L1 trigger signal for its super-cluster.
The generated L1 trigger signal comprises 2 bits, encodes the trigger type, and is typically propagated to the 
FPGA of a so-called cluster service board (CSB).

The hardware design for the L2 stage foresees a single electronics crate 
that fits easily into the body of a Cherenkov camera and can service up
to 2016 pixels. The crate will 
house one L2 Controller Board (L2CB) and 18 CSBs.
Each CSB accepts up to 16 L1 signals and processes them with the help
of a single FPGA. In the simplest case, the CSB produces an OR of the 16 L1 signals and 
the resulting L1$_{\mathrm{CSB}}$ signal is forwarded to a FPGA on the L2CB which is 
generating the L2 camera trigger signal from the 18 L1$_{\mathrm{CSB}}$ signals.

\subsection{FPGA Firmware Design}
\label{sec:fpgas}
\begin{figure}[t]
\centering%
\includegraphics[width=\linewidth]{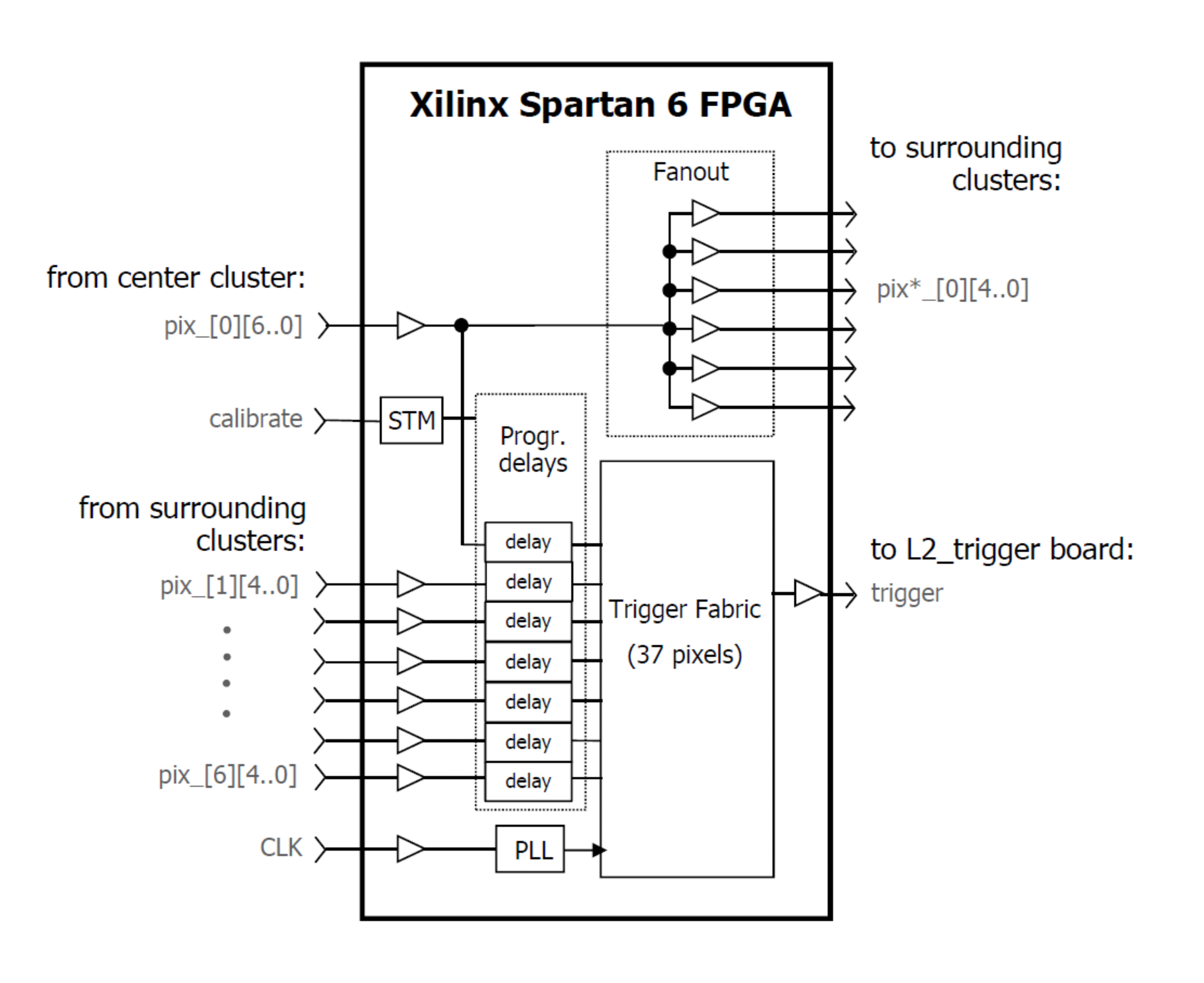}
\caption{Functionality of the Xilinx Spartan 6 FPGA used on the DTB boards. Each FPGA
services one 7-pixel cluster and receives the data of the seven channels (upper left).
These signals are fanned out to up to six neighbouring DTB boards (upper right).
Seven local input signals and 30 signals received from up to six neighbouring
clusters (lower left) pass programmable delays and are processed by a
37-pixel trigger fabric (lower right). Note that only five out of 
the seven signals are transferred to/received from neighbouring FPGAs.
}
\label{fig:fpga}
\end{figure}

FPGAs operate synchronously (i.e.~driven by one or several clocks) 
or asynchronously. Although FPGAs are optimized to run synchronously
it is possible to implement a simple trigger (e.g.~a coincidence of three
neighbouring pixels, three next neighbours (3NN)) asynchronously in 
a purely combinatorial way. A first revision of the DTB board had been equipped with an Altera Cyclone IV FPGA. 
The Altera development system {\em Quartus} allows the very accurate constraining 
of the internal delays. A firmware design implementing a combinatorial 3NN condition 
for 49 pixels showed excellent results. An minimal L0 signal overlap of 1\,ns was sufficient to generate an 
L1 trigger signal.

To enable more complex trigger algorithms that require pipelining the
FPGA has to be used in a synchronous way. When comparing low-cost FPGAs, 
the Xilinx Spartan 6 FPGA has been preferred over Altera's Cyclone IV 
for two reasons. Firstly, any input pin is equipped with an in-system-programmable
delay line for delay adjustments of up to 10\,ns in steps of roughly
40\,ps allowing the finetuning of individual pixel delays. Secondly, 
each pin is equipped with so-called input and output serializer/deserializer
stages (iSer\-Des and oSer\-Des). 
An iSerDes stage converts a serial bit stream of up to 950 million samples per second
(MS/s) into parallel words that can be up to 8 bit wide. The word rate is 
then the serial bit rate divided by the word width, e.g.~800\,MS/s at a word width of 
8 results in a system clock of 100\,MHz. An oSer\-Des stage performs
the reverse operation and converts parallel to serial. 

The second revision of the DTB board described here is based on a Xilinx FPGA of the Spartan 6
family. Figure \ref{fig:fpga} gives an overview of the basic FPGA building blocks
(referred to in {\em italics} in this paragraph) and its inputs and outputs. 
The 7 L0 signals from the corresponding cluster are passed on to six neighbouring 
clusters via fanouts and low-cost flat band cables. They are also fed into a 37-pixel 
{\em Trigger Fabric} which receives the 30 L0 signals from the 6 neighbouring clusters. 
On all I/O channels for L0 signals, {\em Programmable delays} are used
to compensate time differences between the channels. At the inputs of the 
{\em Trigger Fabric}, the iSerDes stages sample the L0 signals
with a rate of 950\,MS/s. With a SerDes factor of 8 this results 
in 8 time slices each of which 1.05\,ns long. It is this sampling of the
L0 signals that provides 1-bit camera images at a rate of about 1\,GHz. 
The {\em Trigger Fabric} comprises, in fact,
eight identical trigger fabrics that work in parallel. Each of the eight trigger
fabrics is connected to one time slice and has 37 input bits. The 8-bit trigger
word resulting from the processing of the trigger fabrics is connected to an 
8-bit oSerDes stage and provides a single serial trigger signal with
a fixed latency. Due to the sampling of the L0 signals the latency fluctuates 
by about $\pm 1$\,ns. By moving the 8-bit words into shift registers the 
history of the L0 signals can be stored and thus be made available for trigger alghorithms 
based on the time distribution of individual pixel signals.

\section{Trigger Prototype Boards and Tests}
\label{sec:prototype}

Hardware implementations of the L0 and L1 stages were developed and 
studied in detail in the laboratory in order to show the validity and 
stability of the hardware concept. The development work was accompanied by extensive 
software tests using a VHDL Test Bench and the Xilinx ISE v13.4 software for design
implementation and simulation. Particular attention has been paid to the
accuracy of timing simulations the results of which have been verified by  
a comparison of FPGA signals (e.g.~trigger patterns) using an oscilloscope.

\begin{figure}[t]
\centering
\includegraphics[width=\linewidth]{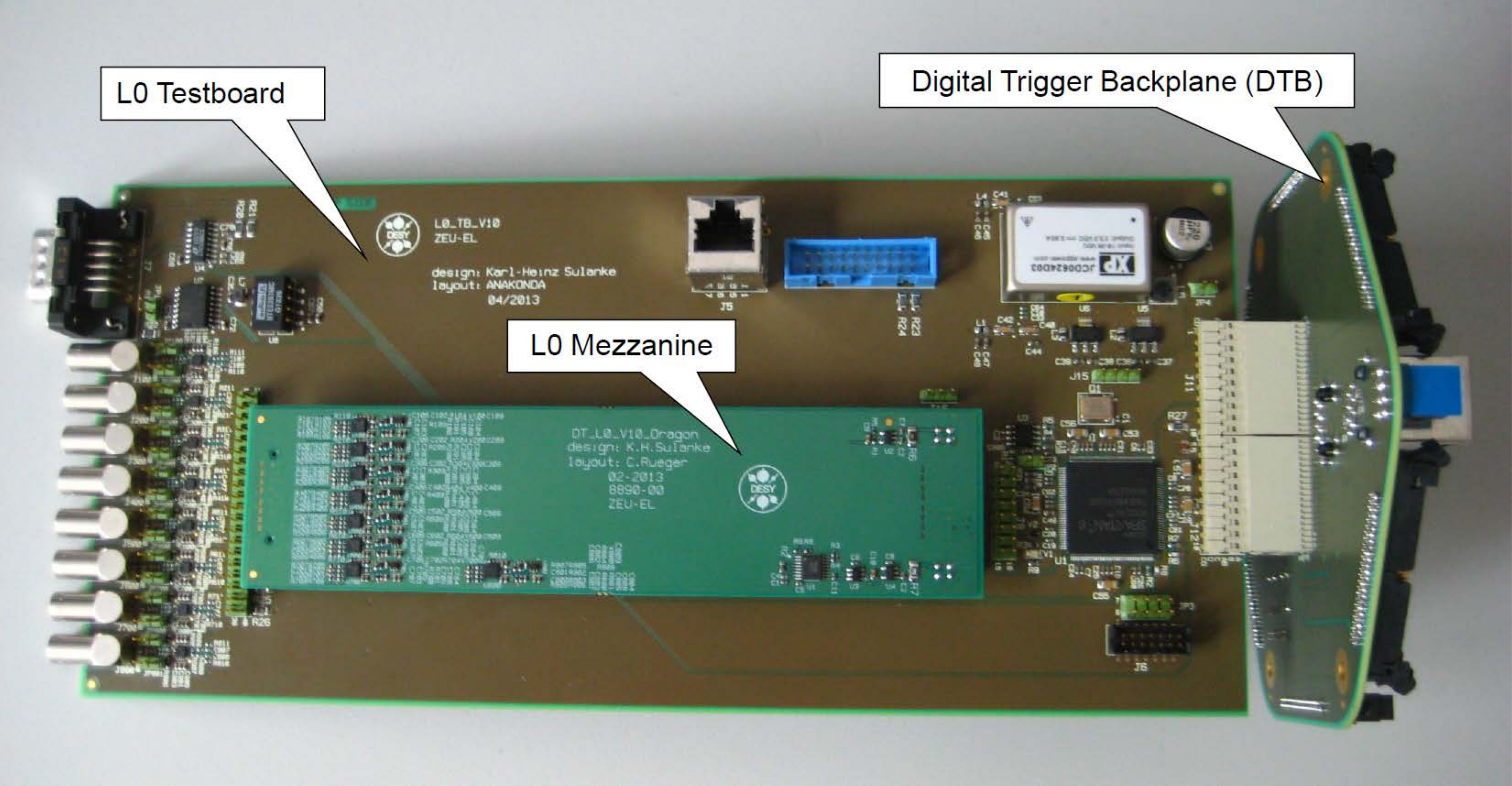}
\vspace{4mm} 
\includegraphics[width=\linewidth]{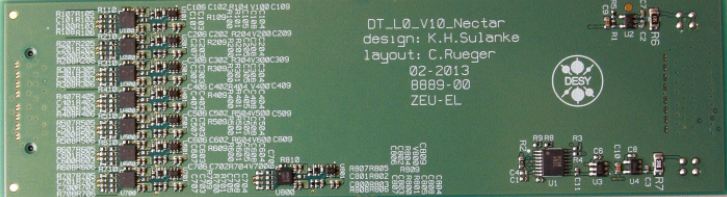}
\caption{
{\em Top:} Assembled FPGA-trigger prototype with a L0 test board
(replacing the FEB in the test setup), L0 mezzanine board and DTB.
The eight lemo connectors on the left-hand side of the FEB can be
used to apply test signals mimicking PMT signals. The FPGA of the
FEB is seen to the right of the L0 mezzanine board. 
{\em Bottom:} Details of L0 mezzanine board. 
}
\label{fig:TrigTestBoard}
\end{figure}

Figure~\ref{fig:TrigTestBoard} (top) shows a test setup
where a dedicated L0 testboard provided the functionality that will come
from the FEB in a real setup. In the following, the L0 testboard will
also be referred to as FEB. 

\begin{figure}[!t]
\centering%
\includegraphics[width=\linewidth]{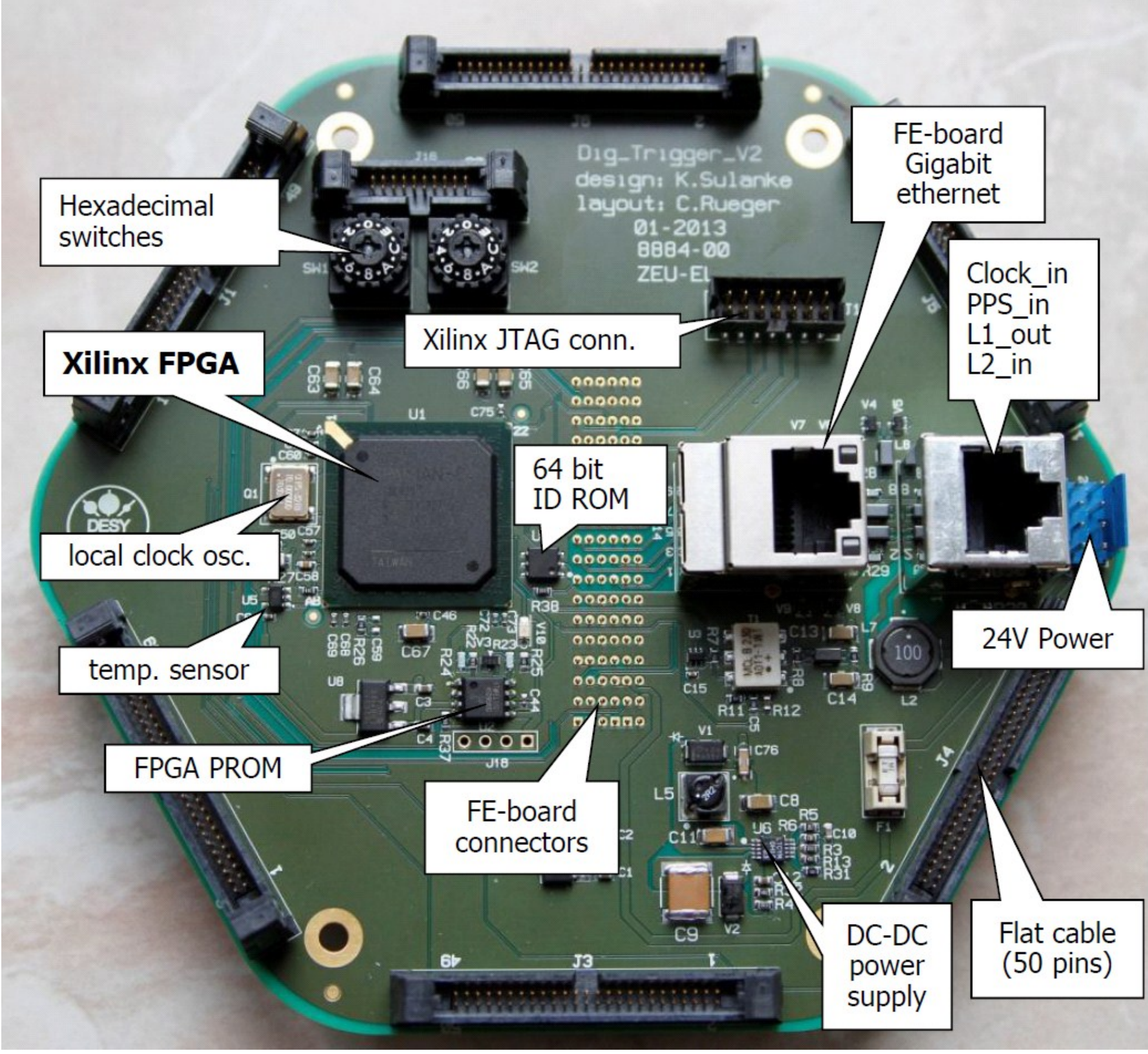}
\caption{Prototype of the Digital Trigger Backplane (DTB) board. 
The main elements are labelled. See text for explanations.}
\label{fig:dtb}
\end{figure}

\subsection{The L0 Mezzanine Board}
\label{sec:l0hardware}
The L0 signals are generated on a mezzanine board (Fig.~\ref{fig:TrigTestBoard}, bottom)
that is mounted on the FEB. This piggy back solution is just
temporary; in its final implementation, the L0 stage will be directly soldered onto the FEB.
The L0 mezzanine board is supplied with $\pm$3.3\,V by the FEB. Linear regulators
generate the $\pm$2.5\,V voltage
supply for the seven analogue input stages. They comprise a low-noise, low-offset amplifier
and a 
fast comparator. 
The amplifier, adjusted to a gain of 5, converts the analogue signal from differential to single ended
and provides in combination with the preamplifier on the FEB a signal of 13\,mV/p.e.~at 
the positive input of the comparator circuit. The negative input is driven
by one output of an eight-channel DAC. The 8-bit DAC is controlled by the FEB FPGA 
via an SPI bus and provides L0 thresholds between 0 and 256\,mV.

\subsection{The Digital Trigger Backplane Board}
\label{sec:l1hardware}
The DTB board is a hexagon-shaped 8-layer printed circuit board (PCB) that is 125\,mm wide. 
An overview of its basic components (referred to in {\em italics} in the next
paragraphs) is shown in Fig.~\ref{fig:dtb}. 
The processing of the 37 L0 input signals is done by a 
low-cost {\em Xilinx FPGA} of the Spartan 6 family of type XC6SLX25$-$2FGG484C. Other parts,
worth mentioning, are a {\em DC-DC power supply}, a {\em local clock oscillator}, two rotating
{\em hexadecimal switches}, a {\em 64-bit ID ROM} and a {\em temperature sensor}.

\begin{figure}[!t]
\centering%
\includegraphics[width=\linewidth]{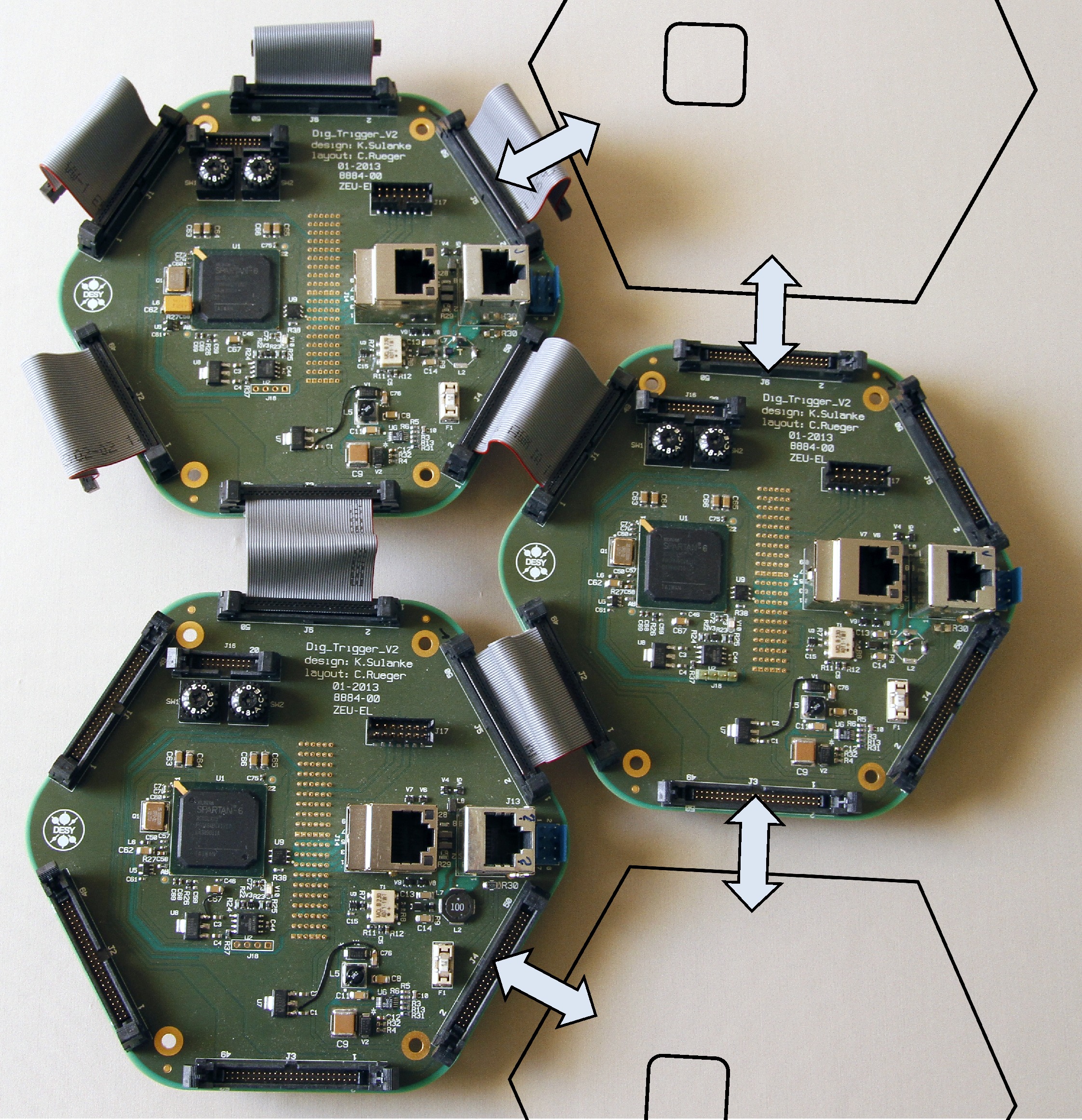}
\caption{Three prototype DTB boards with one Xilinx Spartan 6 cluster FPGA each. 
The flat cables are used for the exchange of L0 signals with the (up to) six
neighbouring DTB boards. See text for more explanations. 
}
\label{fig:TrigHWDemo}
\end{figure}

The seven L0 signals from the central cluster are received from the FEB FPGA 
through the {\em FE-board connectors}. The remaining 30 L0 signals are connected to the 
{\em Xilinx FPGA} using six peripheral 50-pin {\em Flat cable}
connections to the neighbouring clusters (cf.~Fig.~\ref{fig:TrigHWDemo}). The same connections are used to fan-out
the seven L0 signals from the central cluster. After power on, the FPGA gets automatically 
configured by a standard PROM. The 
PROM image can be altered either with the help of an onboard Joint Test Action Group (JTAG) 
connector and a programming cable, or by software from the FPGA on the FEB (exploiting 
the FEB's ethernet connection). The DC-DC power supply circuitry
generates the 3\,V voltage needed by the
two PROMs. Linear regulators
driven by the 3\,V are generating the FPGA's
power supply, 2.5\,V and 1.2\,V. A local clock oscillator is connected to the FPGA. The latter can also be driven
by an external clock. An RJ45 connector is available for the transfer of 
trigger signals ({\em L1\_out}, {\em L2\_in}) and for synchronization by means
of clock signals ({\em Clock\_in}) and/or 1\,Hz-pulses ({\em PPS\_in}).
At the same time this connector can also be used for the 24\,V power connection.
The two {\em FE-board connectors}
(in the centre of the lower 
half of Fig.~\ref{fig:dtb}) establish
the signal, power and {\em FE-board Gigabit ethernet} connection.
Two general purpose rotating hexadecimal switches are useful to provide the FPGA 
with an input representing, e.g.,~the cluster position in the camera.
Additionally, each of the six peripheral 50-pin flat cable connections has a
signal pair for automatic neighbour detection. In this way the same firmware can be used,
independent of the cluster position. The {\em 64 bit ID ROM} assigns a unique ethernet address to 
the FEB. The {\em temperature sensor} may be used to compensate for temperature-dependent 
L0 delay variations.

\subsection{The L2 Crate}
\label{sec:l2hardware0}
The L2 crate with its backplane and its components, the 18 CSBs and one L2CB, are still in the development 
phase. The anticipated size and weight are about 50x20x20$\mathrm{cm}^{3}$ and roughly 11\,kg,
respectively. All boards are powered by the 24\,V that are also used for supplying the 7-pixel clusters. 

\subsection{The Cluster Service Board}
\label{sec:l2hardware1}
\begin{figure}[!t]
\centering%
\includegraphics[width=\linewidth]{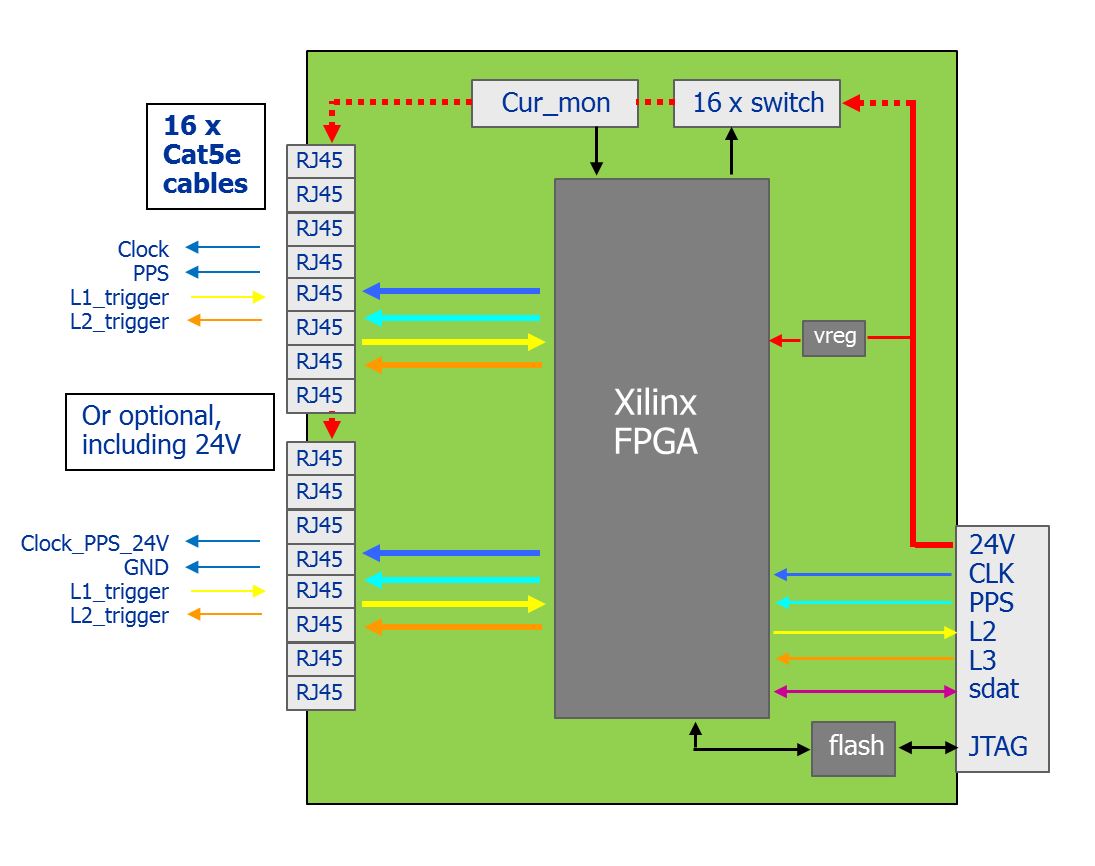}
\caption{Schematic diagram of a L2 Cluster Service Board (CSB). Each CSB communicates 
with 16 cluster FPGAs (via the Cat5e cables and RJ45 connectors shown on the left-hand side).
The L1 signals arriving from the cluster FPGAs are processed by the central
Xilinx FPGA and combined into a L2 trigger signal (for example, by simply ORing all 16
L1 signals). The L2 trigger signal is propagated back to the cluster FPGAs and
to the central L2 Controller Board (cf.~Fig.~\ref{fig:TrigHardware}).
}
\label{fig:CSB}
\end{figure}

The CSB (see Fig.~\ref{fig:CSB}) is still in the design phase. It is based on a low-cost
Xilinx FPGA that should combine arriving L1 signals into a logical OR. On the panel side of the CSB, 
there are 16 RJ45 connectors, one per cluster,
used for distributing a global clock, a PPS and the camera trigger 
signal. The L1 trigger signal gets collected via this channel as well. Optionally, the 24\,V power 
for the cluster can be distributed using the same cable. The board, about 18x16\,cm$^2$ in size, has at 
its back side a direct connector for plugging into the backplane of the L2 crate.

\subsection{The L2 Controller Board}

The L2CB has two interfaces, one pointing outside (away from the camera), 
the other pointing to the inside, establishing the connection to the CSBs. The outside interface 
comprises an Ethernet port and a connection to a central timing unit. The Ethernet connection is 
used to gather slow control information (e.g.~cluster power consumption), for 
cluster-power switching and for configuring the trigger. Event-number driven 
time stamps could be transferred via this channel as well. The interface to a timing unit 
is the connection to a central clock/PPS source. The inside connection, the backplane of the L2 crate, 
carries the clock/PPS signals and the camera trigger. Additionally, there is a bidirectional 
serial communication channel to each CSB for exchanging slow control data and 
trigger configuration data.

\subsection{Hardware Tests}

\begin{figure}[!t]
\centering%
\includegraphics[width=\linewidth]{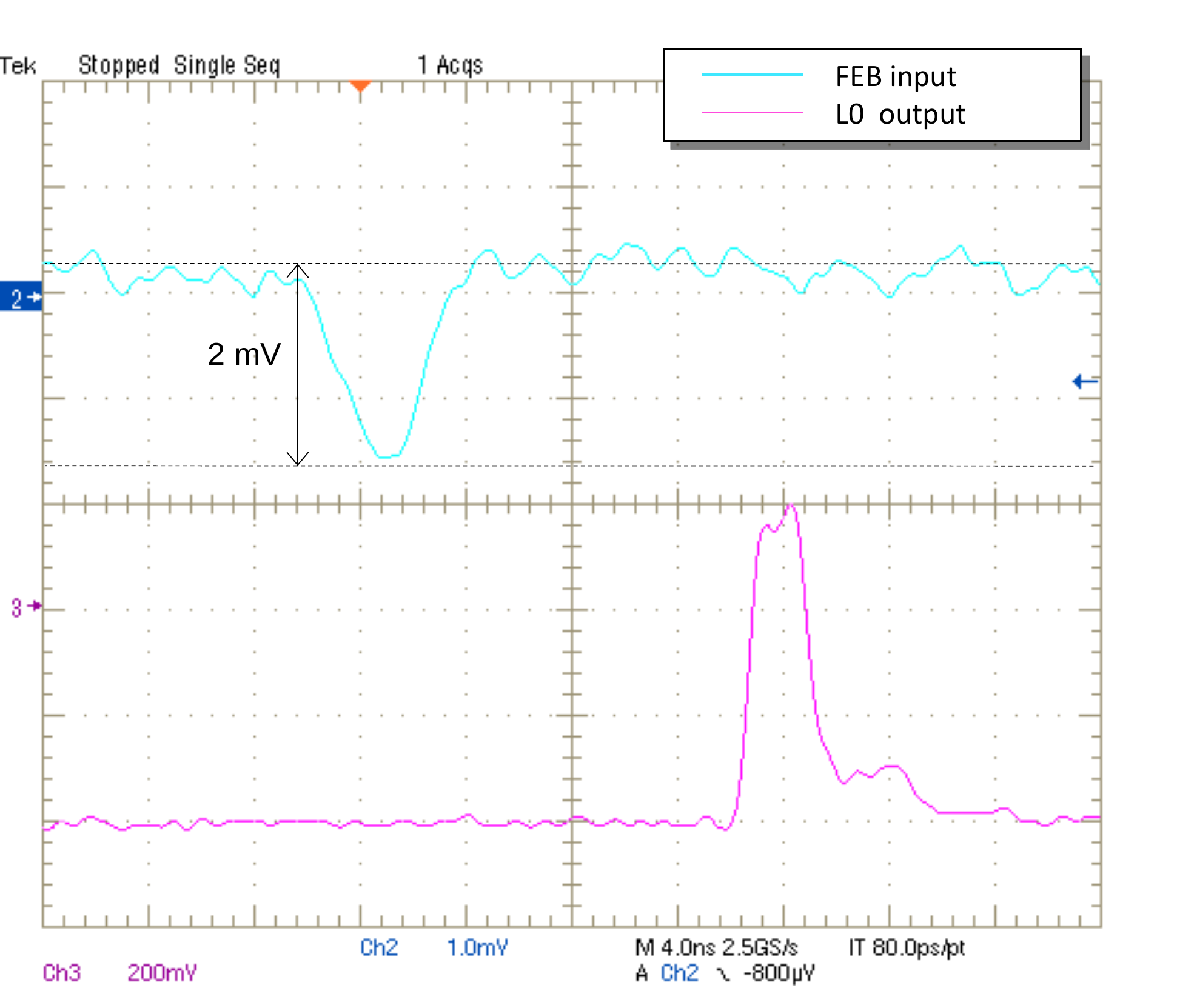}
\caption{
Result of a test of the L0 stage, recorded with a 1\,GHz Tektronix digital
oscilloscope. The blue curve shows the analogue input signal applied to
the FEB. An amplitude of 2\,mV corresponds to 1\,p.e.~The threshold of the corresponding discriminator 
channel was set to 1\,mV. The LVDS L0 signal measured at the discriminator output for such a 
relatively small input signal is shown in red. One box on the horizontal time
axis corresponds to 4\,ns. Note that the time delay between the input and the
L0 signal is dominated by the signal propagation on the cables used in the measurement.
}
\label{fig:TriggerTestL0}
\end{figure}

A number of trigger tests have been carried out using the hardware test bench
described above. The L0 mezzanine board was tested by applying 
generator-driven pulses to the analogue inputs of the FEB.  
The analogue input signal passed an attenuator creating
signal amplitudes as low as 1\,mV. The DAC settings (between 0 and 256\,mV) 
for the adjustment of the discrimination threshold were
controlled by the FEB FPGA via a RS232 (or RS485) interface.
With the selected settings (preamplifier gain 
adjusted to 5) signals with amplitudes down to 2\,mV that correspond to 1\,p.e.~were securely detectable
(as shown in Fig.~\ref{fig:TriggerTestL0}). Channel-to-channel skews were found to be
smaller than 50\,ps; a suppression of more than 55\,dB was measured for the amplitude of 
analogue crosstalk signals for neighbouring channels.

\begin{figure}[!t]
\centering%
\includegraphics[width=\linewidth]{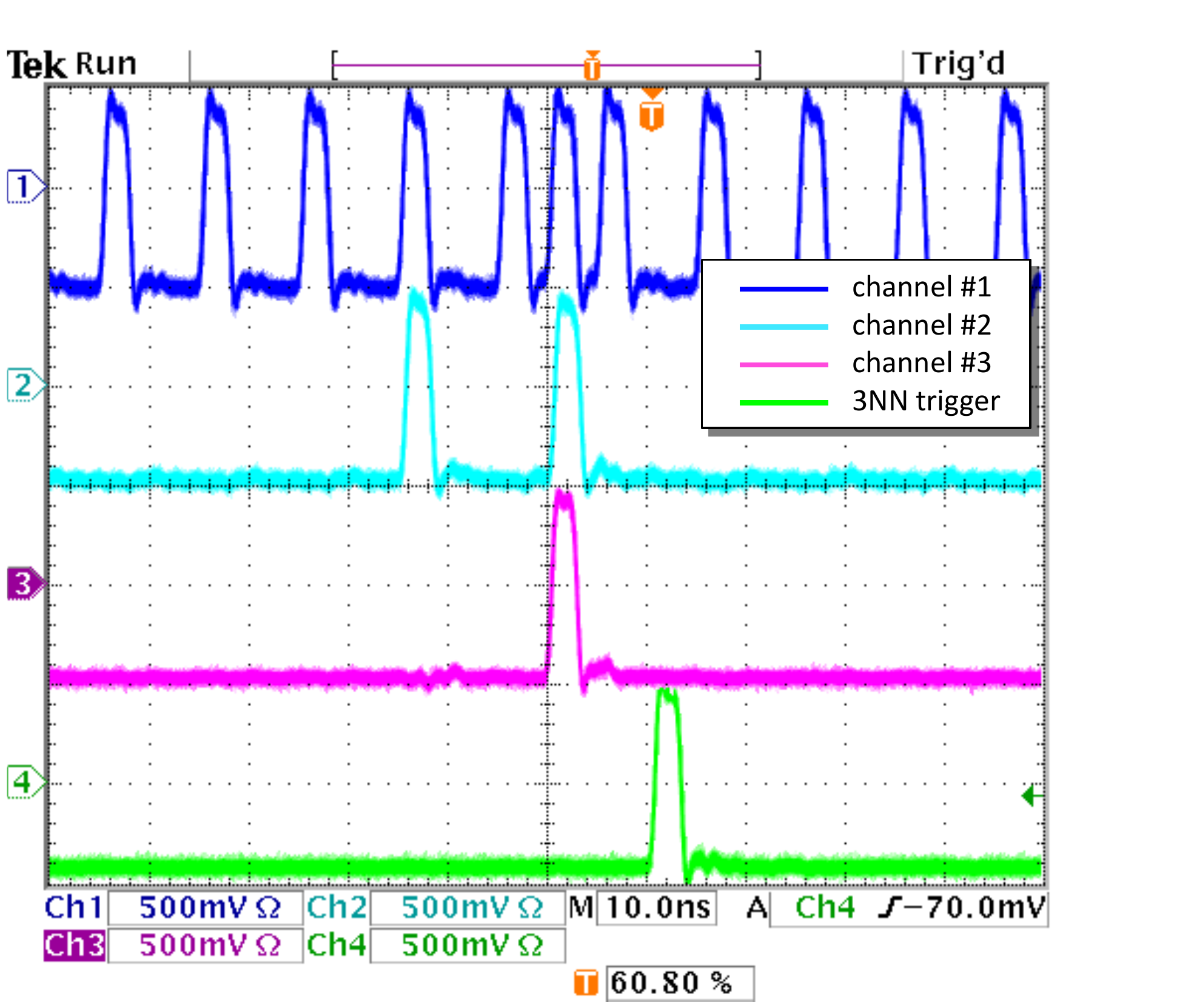} 
\caption{Test of a L1 stage with a majority trigger scenario for three neighbouring
channels (three next neighbours, 3NN), recorded with the same equipment as used for 
Fig.~\ref{fig:TriggerTestL0}. The trigger response to a generated 3NN pattern 
(channels 1, 2, and 3), shown in green, was generated using an asynchronous 
firmware design. The time scale is 10\,ns per box.
}
\label{fig:TriggerTest1}
\end{figure}

An efficient way to test the L1 stage, the DTB, is the connection of two
DTB boards. In this setup, one board acts as pattern generator,
while the second board functions as trigger board.
The pattern generator, simply a particular firmware version, is based on
a look-up table of 38\,bit x 8k (up to 37k). 37 bits were used to
emulate the 37 L0 signals, while the 38th bit decoded the expected outcome
of the trigger. Figure~\ref{fig:TriggerTest1} shows an example of such a test
with a first revision of the DTB boards (equipped with Altera Cyclone IV FPGAs).
The first three curves (blue and magenta) correspond to the L0 signals
on three neighbouring channels and one sees the generated L1 signal (green) 
that occurred when the FPGA had been programmed to accept coincidences
of three neighbouring channels (three next neighbours, 3NN) in an
asynchronous firmware design. 

\subsection{Power Consumption and Price}

For completeness, estimates of the power consumption and the price
of the digital trigger are included here. Both power consumption and
price are dominated by the DTB boards; the estimated power consumption 
for the L2 crate is 50\,W.

A cost estimate of the various components 
on a DTB board can be found in Tab.~\ref{tab:price}.
The listed prices are based on the assumption that more than 1000 DTB
boards will pro produced. The estimated price per pixel is 19~\euro, which is 
about 12\,\% of the targeted channel cost (150--160~\euro). The power consumption 
of the DTB board is dominated by the FPGA which uses between 1.5 and 2\,W
depending on the complexity of the executed trigger algorithm. The 
generation of the L0 signals takes about 0.2\,W per pixel, so the
overall power consumption is between 0.4\,W and 0.5\,W per pixel.
This power consumption is quite similar to the overall power consumption of 
analogue camera trigger designs \cite{AnalogTrigger}.

\begin{table}
\caption{\label{tab:price}
Price estimates for the components of one DTB board serving 
as trigger board for seven PMTs. Price reductions arising from the production 
of more than 1000 trigger boards have been assumed.
}
\vspace{0.2cm}
\centering
\begin{tabular}{p{0.75\linewidth}r}
\hline
component & price \\ 
\hline
trigger PCB & 46 \eur \\
FPGA & 17 \eur \\
connectors and cabling & 50 \eur \\
board assembly & 20 \eur\\
\hline
total & 133 \eur \\
total, per pixel & 19 \eur \\
\hline
\end{tabular}
\end{table}

\section{Simulated Performance}
\label{sec:simu}

In parallel to the design and test of trigger hardware, the
performance of FPGA-based camera trigger algorithms has been
investigated with the help of Monte Carlo simulations. A full comparison
of the investigated trigger algorithms, the dependence of their
performance on the NSB level, the telescope type, and the bandwidth of the used 
electronics are beyond the scope of this paper and are   
the subject of earlier (\cite{TrigSim2011,TrigSim2013}) and present work.
The simulation results presented in the following are restricted
to the case of Davies-Cotton type MSTs with a mirror area 100\,m$^2$ and 
a focal length of 15.6\,m that are equipped with 1765-pixel PMT cameras \cite{CtaMc}
with a high-bandwidth front-end electronics and an FPGA-based trigger operating
in asynchronous mode. The actual simulation of digital camera trigger algorithms followed 
closely the capabilities of a design (cf.~Sec.~\ref{sec:fpgas}) based on 
Altera FPGAs operated at 300\,MHz in asynchronous mode. The conducted Monte Carlo
studies had two objectives, namely (i) to show the importance of selecting a proper
camera trigger algorithm and (ii) to illustrate how the combination of different trigger algorithms (which is easily 
possible with the FPGA-based trigger described here, but impossible with hard-wired trigger schemes) can improve
the performance of a camera trigger. For simplicity and in contrast to the detailed design 
presented in Section~\ref{sec:digi} the simulation assumed that the
FPGAs have access to all 49 pixels of a super-cluster (and not only to 37 pixels).
Given the high degree of overlap between super-clusters the differences
between the 49-pixel version and the 37-pixel version should be
small.

\subsection{Camera Trigger Simulations}

\begin{table}[t]
\caption{\label{tab:refelec}
Summary of the PMT properties and the NSB levels applied in the simulation
of camera trigger algorithms. In connection with the excess noise factor, $F$ (see, e.g., \cite{hamamatsu}) is defined as
$F=1+v/\mu^2$ where $\mu$ ($v$) denotes the mean (variance) of the amplitude distribution
of single-pe pulses. The after-pulse probability refers to signals with 
$\ge 4\,\mathrm{p.e}$. The NSB level is the rate of photoelectrons generated
at the PMT cathode by the night-sky background.
}
\vspace{0.2cm}
\centering
\begin{tabular}{lr}
\hline
parameter & value \\
\hline
 PMT signal shape & Gaussian (FWHM=2.6\,ns) \\
 PMT jitter & 1.5\,ns (FWHM)\\
 PMT excess noise factor $\sqrt{F}$ & 1.175 \\ 
 PMT after-pulse probability & 0.02\,\% \\
 NSB level & 104--467\,MHz  \\
\hline
\end{tabular}
\end{table}

The simulations of various camera trigger algorithms were
performed with the \trigsim\ software package (see \cite{TrigSim2011}
and \ref{sec:trigsim}). Simulated air showers from protons and 
$\gamma$-rays at a zenith angle of $20^{\circ}$ were processed with the CTA detector simulation program 
\simtel\ \cite{simtel} that provides the arrival times of shower-induced 
photoelectrons in the cameras of the Cherenkov telescopes. These photoelectrons were
then injected into \trigsim\ in order to carefully examine various trigger algorithms
and readout options.

\begin{figure}[t]
\centering%
\includegraphics[width=\linewidth]{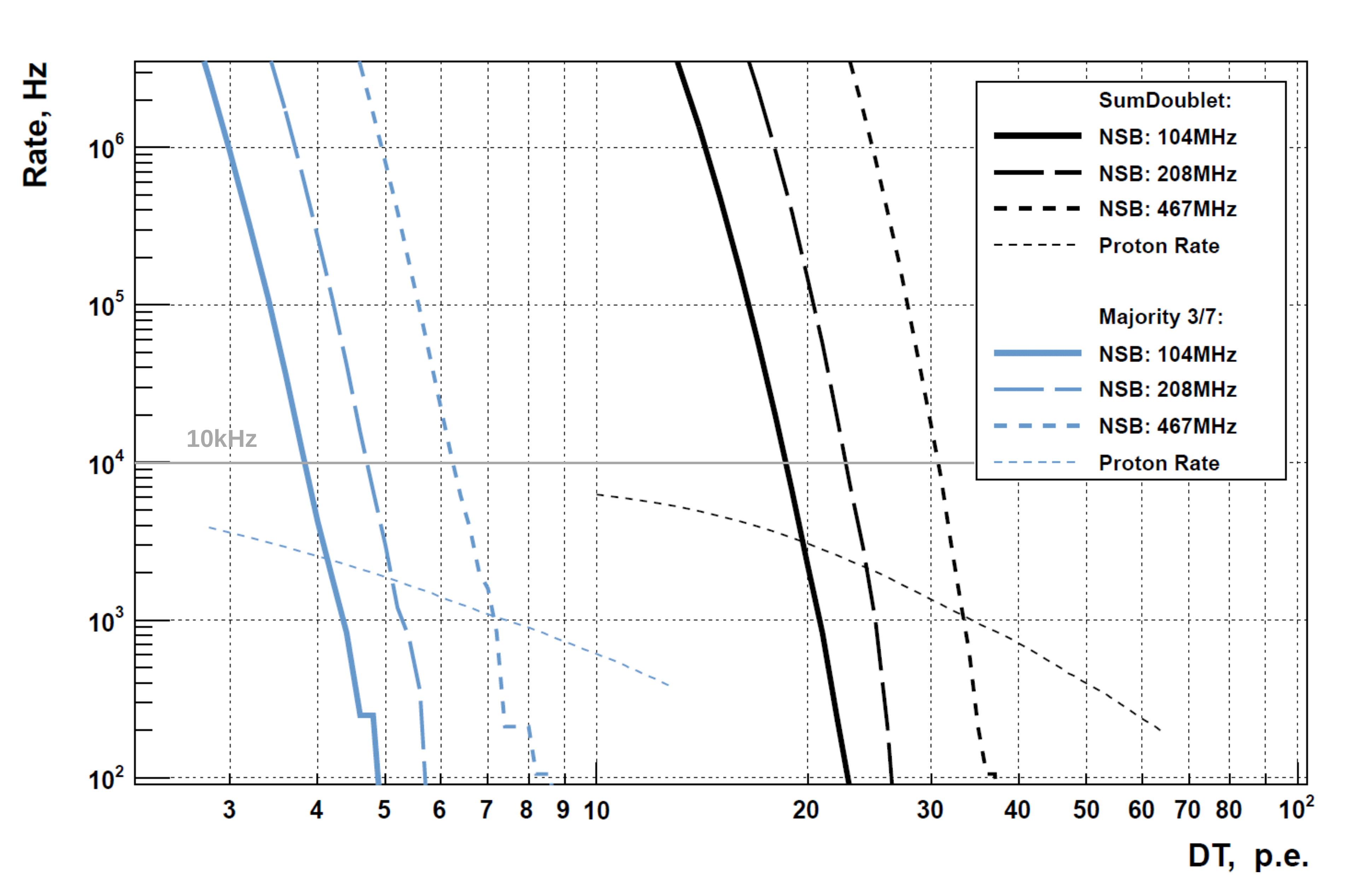}  
\caption{Accidental MST camera trigger rate versus discriminator threshold for different 
NSB levels (104\,MHz, 208\,MHz, 467\,MHz) and two different trigger
algorithms (SumDoublet and Majority 3/7). The trigger algorithms and their parameters are explained in 
detail in the text. The horizontal line marks the chosen working point
(NSB-induced trigger rate of 10\,kHz); the two less steep curves (short-dashed) denote the 
total camera trigger rate expected from cosmic-ray proton showers.
}
\label{fig:nsb}
\end{figure}

In the \trigsim\ simulation, the time development of the PMT signal in every camera 
pixel was simulated in a window of up to 200\,ns length with a time resolution of 
$0.2$\,ns. The signal shaping implied by the PMT and the associated electronics was 
applied to each single p.e.~and the shaped p.e.~pulses were summed to obtain the full signal. Table
\ref{tab:refelec} summarizes the signal shaping (a Gaussian with a FWHM of 2.6\,ns) and the NSB rates 
that were used for this study. Photosensor after-pulses were simulated at the 
level of 0.02\,\% above 4 photoelectrons. The resulting 'analogue camera image' for an event
was then subjected to a number of different analogue and digital trigger algorithms
in time and space. This procedure enables a strict unbiased comparison since 
statistical fluctuations in the analogue camera images impact all trigger algorithms in the
same way.

To determine accidental trigger rates due to the NSB-induced pulses and their after-pulses the response of the camera 
to continuous light with various intensities was simulated. Figure~\ref{fig:nsb}
shows the accidental trigger rate of an MST camera for 
NSB levels of 104\,MHz (solid), 208\,MHz (long dashes), and 467\,MHz (short dashes) that roughly correspond
to the observational conditions expected for CTA. The two lower NSB levels refer to observations of
average extra-galactic and galactic field of views;  the highest NSB level (4.5 times the extra-galactic
NSB) stands for extreme observation conditions (e.g.~during moonshine). The rates are shown
as function of the discriminator threshold value (labelled "DT''), calibrated in units 
of photoelectrons collected at the last PMT dynode. Different
camera trigger algorithms (detailed below) are shown in black and blue,
respectively. For each algorithm the trigger threshold DT was chosen such that 
an accidental camera trigger rate of 10\,kHz occured. With this
setting, the collection area for $\gamma$-rays is a simple figure of merit for the 
different camera trigger algorithms. 

\subsection{Camera Trigger Algorithms}

One analogue and several digital camera trigger algorithms have been studied in detail in order
to explore a variety of options. The parameters of all algorithms 
(e.g.~coincidence windows, clipping levels) were optimized to
obtain the best performance at a given NSB suppression. An analogue sum trigger 
is not possible with the hardware described above but is generally considered as the most suitable 
solution for the LSTs operating at the lower end of the CTA energy range.  The 
simulation of an analogue sum trigger was therefore included as a reference trigger:
\begin{description}
\item[SumDoublet] The simulated analogue sum trigger is based on
    overlapping trigger patches that comprise 14 pixels in two neighbouring
    7-pixel clusters. Such a trigger patch is referred to as {\em doublet}
    and there are 12 overlapping doublets in a super cluster (cf.~the inset in Fig.~\ref{fig:CameraClusters}).
    Pulses were clipped at the level of 7.5 p.e.~and the threshold 
    of the trigger patches was 18 p.e. (for the default NSB level of 104\,MHz). 
\end{description}
In the simulation of digital camera trigger algorithms executed on an FPGA in asynchronous mode, 
the minimum overlap of the L0 signals that is required to produce a logical AND in an FPGA was
set to $\tau=1$\,ns. The coincidence window is then defined
by $\tau$ and the length of the L0 signal. The effective coincidence window was optimized 
by varying the L0 signal length in steps of 3.3\,ns. The explored algorithms were:
\begin{description}
\item[Majority 3/7] This digital majority trigger required $N_{\mathrm{maj}}=3$ 
    out of the 7 pixels in a trigger patch to be above the pixel threshold DT. 
    The trigger patches were defined around those 25 pixels in a 
    super cluster that are not located at the super cluster boundary 
    (cf.~the inset in Fig.~\ref{fig:CameraClusters}); 
    the trigger patch comprised the central pixel and its six neighbouring pixels. 
    The L0 signal length was 3.3\,ns and a coincidence of all $N_{\mathrm{maj}}$ pixels was required.
    The pixel threshold was 3.8 p.e. at 104\,MHz. 
\item[Majority 4/7] Like Majority 3/7 but with $N_{\mathrm{maj}}=4$ 
    and correspondingly lower pixel threshold.
\item[Binary Trigger: 3/7 or 4/7] This algorithm is the logical OR
  of refined versions of the Majority 3/7 and Majority 4/7 algorithms that were assumed to be
  executed in parallel in the FPGA trigger fabrics. The L0 signal length was 3.3\,ns. Each of the majority
  triggers had an independent pixel threshold (4.0 and 3.6 p.e. at 104\,MHz, respectively)
  and a coincidence was only required for neighbouring pixels (and 
  not for all pixels like in the case of the Majority 3/7 and Majority 4/7 triggers).
  Only coincidences occurring in a time window of less than 25\,ns 
  contributed to the trigger.
\item[Majority 5/21] This majority trigger required $N_{\mathrm{maj}}=5$ 
     pixels above a threshold of 3.0 p.e.~at 104\,MHz in a trigger patch 
     comprising three clusters (21 pixels). The trigger patches comprised
     the central cluster of a super cluster and two clusters at the super cluster
     boundary that are adjacent to each other. There are six such overlapping
     trigger patches. The L0 signal length was 6.6\,ns.
\item[Majority 7/21] Like Majority 5/21 but with $N_{\mathrm{maj}}=7$, 
    a lower pixel threshold (2.8 p.e.~at 104\,MHz), and an L0 signal length of
    9.9\,ns.
\end{description}
It is noted here that an actual implementation of the Binary Trigger would require
two comparators for L0 signals above independent pixel thresholds. This feature
is not present in the prototype hardware described in Section~\ref{sec:prototype} but could
be added in a future version.

\subsection{Comparison of Camera Trigger Algorithms}

\begin{figure}[t]
\centering%
\includegraphics[width=\linewidth]{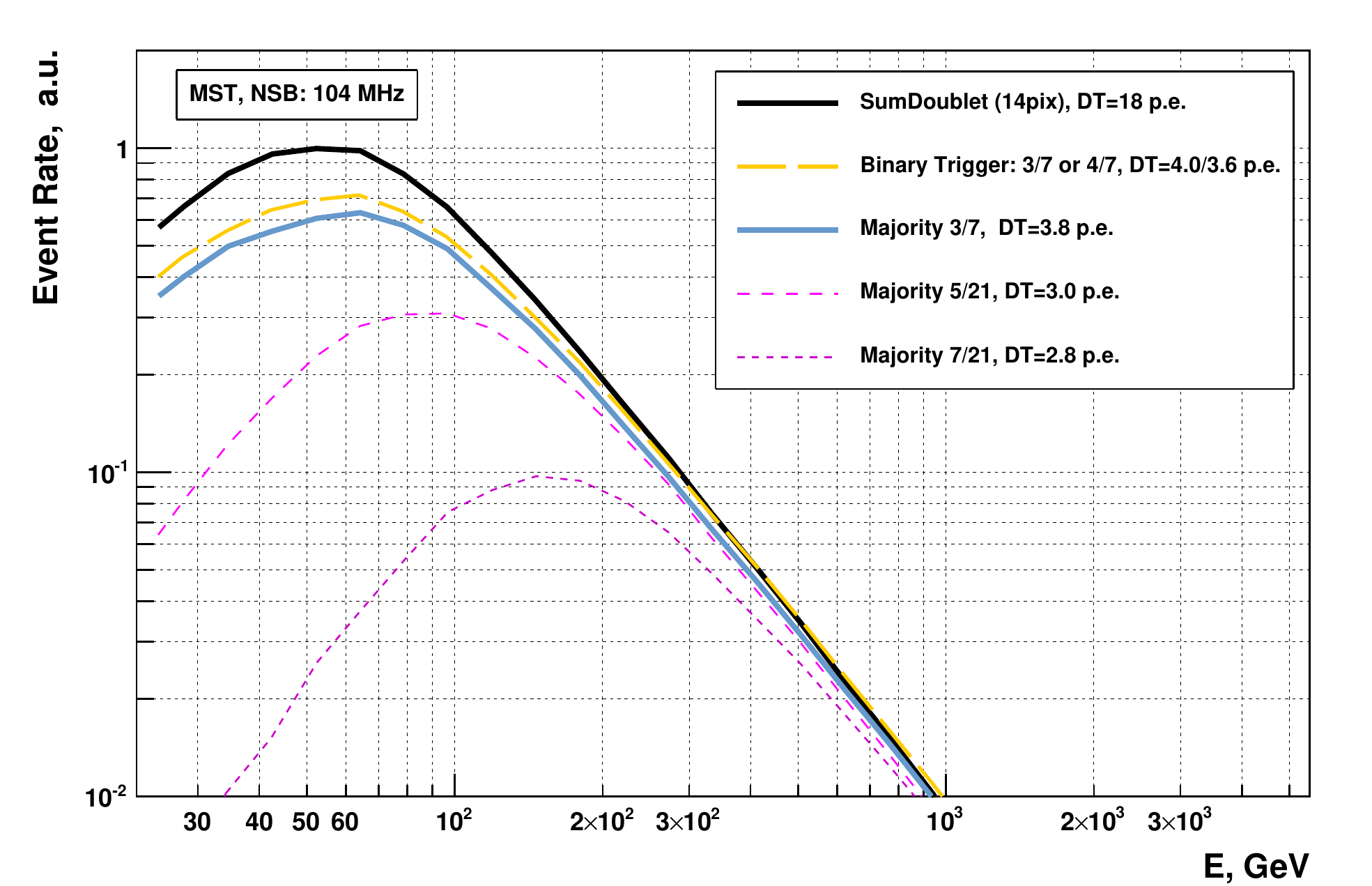}  
\caption{Normalized trigger rates as a function of the simulated $\gamma$-ray
energy for an MST operated with different FPGA-based camera triggers. 
The simulated NSB level was 104\,MHz, and the thresholds and parameters of all algorithms 
had been adjusted such that the NSB-induced camera trigger rate is 10\,kHz. 
The different trigger algorithms are described in the text. The differential trigger rates have been
calculated for a gamma-ray source with a photon index of 2.45. All rates
have been normalized such that the rate assumes a value of unity at the analogue sum trigger threshold of 55\,GeV.
}
\label{fig:rates}
\end{figure}

Figure~\ref{fig:rates} shows the normalized trigger rates of the discussed camera trigger
algorithms as a function of the simulated $\gamma$-ray energy. The simulated NSB level was 104\,MHz and 
the NSB-induced camera trigger rate was fixed to 10\,kHz. The trigger rates have been
calculated for a $\gamma$-ray source with a photon index of 2.45. The energy threshold of
each algorithm (defined as the energy where the maximum trigger rate occurs) was estimated 
with a typical error of 5\,\%. All rates have been normalized such that the rate assumes a 
value of unity at the threshold of the SumDoublet trigger (black solid line).

\begin{figure}[t]
\centering%
\includegraphics[width=\linewidth]{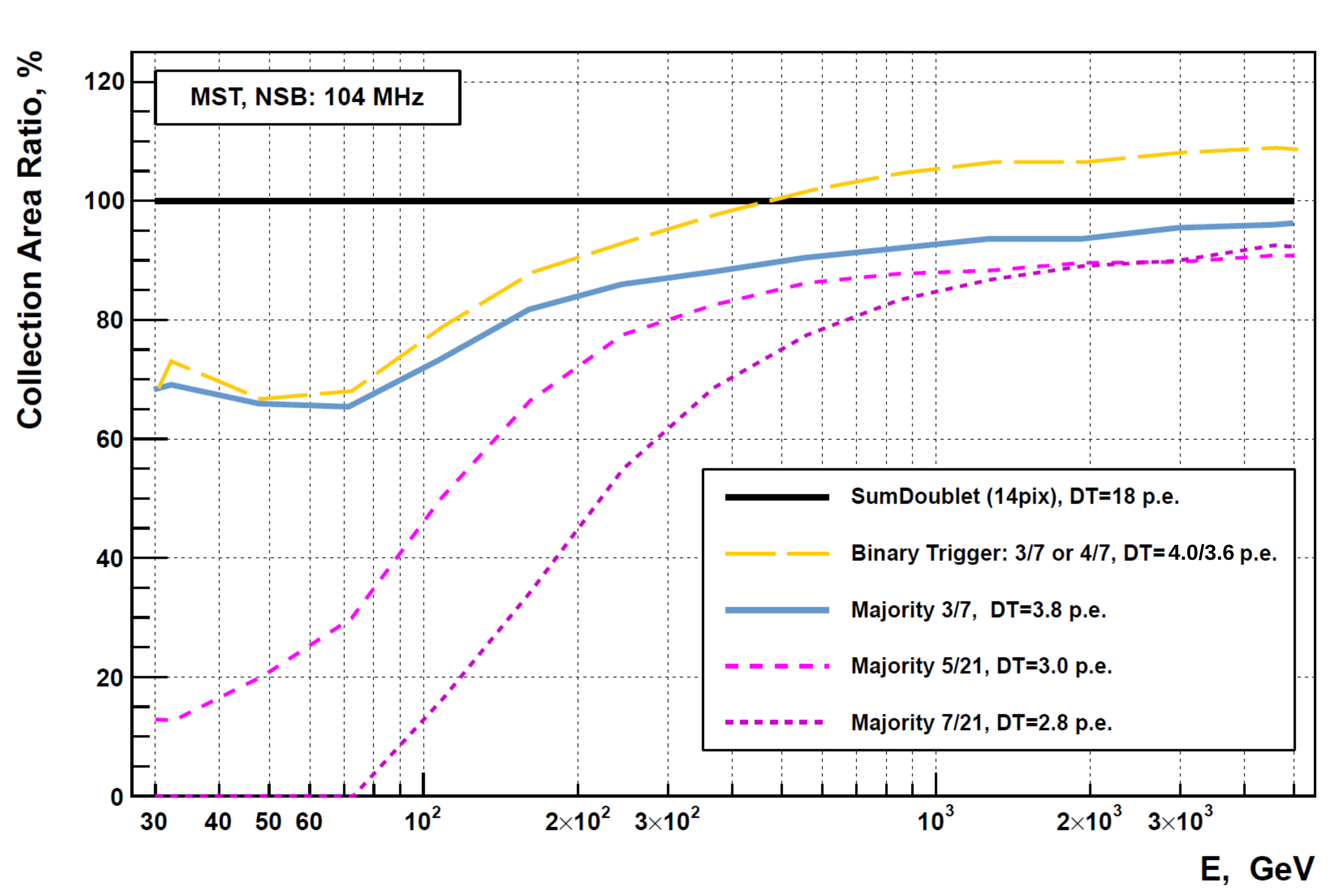}  
\caption{Normalized collection areas as a function of the simulated $\gamma$-ray 
energy for an MST operated with different FPGA-based camera triggers. The simulation
settings were like in Fig.~\ref{fig:rates}. The effective areas have been normalized to the effective 
area of the analogue sum trigger. 
}
\label{fig:collarea}
\end{figure}

In in a similar way, the relative $\gamma$-ray collection area as a function of simulated $\gamma$-ray 
energy is depicted in Fig.~\ref{fig:collarea} for the same simulation settings. 
All curves have been normalised to the SumDoublet trigger the good performance of which has been 
used as a reference. It is evident from Figs.~\ref{fig:rates} and \ref{fig:collarea} that the analogue 
SumDoublet trigger (black lines) provides the lowest threshold (55\,GeV) and the 
largest collection area below 500\,GeV making it the algorithm of choice for
telescopes operating at the lower end of the CTA energy range. The Majority 5/21 and 
Majority 7/21 triggers (magenta, short-dashed curves) have poor
performance for low photon energies since the 21-pixel trigger patches 
require a rather high pixel multiplicity ($N_{\mathrm{maj}}=5$ or 7) to keep the NSB rate 
under control. At such high pixel multiplicities, many small $\gamma$-ray shower
images will not trigger the camera. The resulting thresholds are 90\,GeV and 
150\,GeV for the Majority 5/21 and Majority 7/21 algorithms, respectively. 
The Majority 3/7 trigger (blue curves) performs better than 
a Majority 4/7 trigger (not shown in Figs.~\ref{fig:rates} and \ref{fig:collarea}) and was found to be the 
best choice among all simple majority triggers. Its threshold is 70\,GeV and its collection area for 
photons at the typical MST threshold of about 100\,GeV is only 30\,\% lower 
than for the SumDoublet trigger; this difference diminishes to 
10\,\% at energies greater than 1\,TeV. The Binary
Trigger: 3/7 or 4/7 (long-dashed, yellow), if implemented, would have the same threshold 
but would be more performant than the 
simple Majority 3/7 trigger at all energies. In the simulation, their collection areas
below the MST threshold are comparable but still 30\,\% lower than the
collection area obtained with an analogue sum trigger. The effective area
of the binary trigger converges, however, quicker towards the collection 
area of the SumDoublet trigger and surpasses its performance 
at energies greater than 500\,GeV. The binary trigger also accepts
high-energy showers with large impact parameters and hence 
sizable time gradients since a coincidence in time is only required for
adjacent pixels. Overall, the performance of the binary trigger demonstrates clearly how the paralled 
execution of trigger algorithms with the help of low-cost FPGAs can help to
exceed the performance of other approaches.

\section{Summary}
\label{sec:summary}

The presented design of an FPGA-based digital trigger allows 
the triggering of cameras with an estimated latency between 80--300\,ns, depending
on complexity of the executed algorithm. The design grants the trigger algorithms
access to overlapping camera regions that are large enough to contain 
a sizable fraction of a typical shower image. Each trigger algorithm
can exploit the information for 37 pixels in its region in the space and time domain. 
Several trigger algorithms can be executed in parallel, and their
results can be combined to result in a more informed trigger decision. This allows,
for example, the deployment of different algorithms that are focussed on the lowest and highest 
$\gamma$-ray energies, respectively, and whose results can be combined to obtain
a superior performance over the full energy range. 
Thanks to the use of re-programmable FPGAs, an update of the trigger algorithms is 
quickly possible, thus the most suitable algorithm for certain observation conditions 
and the targeted $\gamma$-ray source can be chosen. In addition to the execution of
trigger algorithms the FPGAs could provide input for the acquisition and online reduction of
the pixel data (adjustment of the readout window in time, zero suppression).
This flexibility and extensibility are inherent features of this trigger design and come at no extra effort. 

The functioning of the produced prototype boards and the successful
tests of the different trigger stages suggest that the digital camera 
trigger is viable option for Cherenkov cameras in CTA. The power 
consumption and the estimated price per channel are well within budget of a Cherenkov camera. 

\section{Acknowledgements}
The authors would like to thank the members of the CTA consortium for stimulating discussions.
We are grateful to the anonymous referee whose comments helped to improve the manuscript.
Konrad Bernl\"ohr is acknowledged for help with the CTA simulation tools and for
comments on Sec.~\ref{sec:simu}. We also would like to thank Oscar Blanch Bigas for 
his valuable contributions to \trigsim.

\appendix

\section{The \trigsim\ Simulation}
\label{sec:trigsim}
The \trigsim\ software package was developed within the Monte Carlo
simulation effort for CTA in order to quickly explore the large
variety of camera trigger algorithms and their parameters. It is 
available on request from the authors\footnote{Contact U.~Schwanke (schwanke@physik.hu-berlin.de)}
and consists of a library (written in C$^{++}$) and a main
program that is also called \trigsimp. The package depends
on two external software packages, namely {\tt root} \cite{root} and 
{\tt hessio}. The {\tt hessio} package implements the
platform-independent {\tt eventio} \cite{simtel} format that is used to store the output of CTA
detector simulations in files.

The input to \trigsimp\ are {\tt eventio} files with the output of the detector
simulation program \simtel\ \cite{simtel} that simulates the
response of CTA candidate arrays to the Cherenkov light emitted 
by air showers initiated by photons, electrons and hadrons. The input
files contain the geometry of the detector (arrangement of telescopes,
camera geometry), the simulated particles (type, energy, direction),
and for each camera pixel in a telescope the times when single
photoelectrons were created at the PMT cathode\footnote{Storage of photoelectrons
is enabled by setting {\tt SAVE\_PHOTONS=2} in \simtel}. \trigsim\
copies the detector geometry from the input files and can not, of course, change the PMT 
efficiency a posteriori, but all other aspects of the
signal processing (NSB rates, PMT jitter, signal shaping, digitization etc.) 
can be changed under the control of configuration files and command
line arguments. For most applications \trigsimp\ loops over the 
events in the input files, but it can also be steered to disregard
the p.e. information and to just simulate the response to NSB photons
that are created internally. In the latter case, the input files solely serve
to extract the detector geometry. The simulation output of \trigsimp\ is
typically stored in {\tt root} files that can be directly analysed or
fed into a shower reconstruction program. For debugging and cross-checks
\trigsimp\ allows the creation of graphical displays of camera images
and of simulated signals; it is also possible to inspect individual events
while the program waits for a keyboard input. 

The central engine of \trigsim\ is a base class ({\tt Trigger::TriggerBase}) 
defining a set of virtual member functions that accept data blocks as 
arguments (containing, for example, the detector geometry or the photoelectrons for a single shower) and
which are called by the \trigsimp\ main program when reading Monte Carlo input files. 
For a typical simulation study one has to subclass the base class (or one
of the already existing derived classes), overwrite the virtual functions as desired,
and modify a static function that returns an instance of the new class
to the \trigsimp\ main program; there is no need to modify the main
program and one can switch from one simulation study to another by passing
a command line argument to \trigsimp.




\begin{thebibliography}{00}

\bibitem{CtaConceptShort} Acharya, B.~S. et al. [CTA Consortium], \emph{Astroparticle Physics} {\bf 43} (2013) 3.

\bibitem{CtaMc} Bernl\"ohr, K.~et al. [CTA Consortium], \emph{Astroparticle Physics} {\bf 43} (2013) 171.

\bibitem{timegradient} {He{\ss}}, M.~et al. [HEGRA Collaboration], \emph{Astroparticle Physics} {\bf 11} (1999) 363.

\bibitem{HessArrayTrigger2004} {Funk}, S.~et al., \emph{Astroparticle Physics} {\bf 22} (2004) 285.

\bibitem{German2008} {Hermann}, G.~et al., \emph{American Institute of Physics Conference Series} {\bf 1085} (2008) 898.

\bibitem{Krennrich2009} {Schroedter}, M.~et al., arXiv:0908.0179, 2009.

\bibitem{colibri} Naumann, C.~L.~et al., \emph{Journal of Instrumentation} {\bf 8} (2013) P06011.

\bibitem{MagicTrigger2004} {Meucci}, M.~et al., \emph{Nuclear Instruments and Methods in Physics Research A} {\bf 518} (2004) 554.

\bibitem{zitzer2013} {Zitzer}, B.~for the VERITAS Collaboration, Proceedings of the \emph{International Cosmic Ray Conference 2013} Rio de Janeiro, arXiv:1307.8360, 2013.

\bibitem{HessTrigger2003} Vincent, P.~et al. [H.E.S.S.~Collaboration], \emph{International Cosmic Ray Conference} {\bf 5} (2003) 2887.

\bibitem{MagicSumTrigger2008} {Aliu}, E.~et al. [MAGIC Collaboration], \emph{Science} {\bf 322} (2000) 1221.

\bibitem{AnalogTrigger} Barcelo, M.~et~al.~for the CTA Consortium, arXiv:1307.3169, 2013.

\bibitem{TrigSim2011} Wischnewski, R., Schwanke, U., Shayduk, M.~and Sulanke, K.~for the CTA~Consortium, Proceedings of the \emph{International Cosmic Ray Conference 2011}, Beijing, vol.~{\bf 9}, p.~63, arXiv:1111.2183.

\bibitem{TrigSim2013} {Shayduk}, M., Vorobiov, S., {Schwanke}, U.~and Wischnewski, R.~for the CTA Consortium, Proceedings of the \emph{International 
Cosmic Ray Conference 2013}, Rio de Janeiro, arXiv:1307.2232, 2013.

\bibitem{simtel} Bernl{\"o}hr, K.~\emph{Astroparticle Physics} {\bf 30} (2008) 149.

\bibitem{hamamatsu} https://www.hamamatsu.com/resources/pdf/etd/PMT\_handbook\_v3aE.pdf

\bibitem{root} Brun, R.~and Rademakers, F. \emph{Nuclear Instruments and Methods in Physics Research A} {\bf 389}, (1997) 81.

\end{thebibliography}



\end{document}